\documentclass[aip,amsmath,amssymb,reprint]{revtex4-1}

\usepackage{graphicx}
\usepackage{bm}

\usepackage[utf8]{inputenc}
\usepackage[T1]{fontenc}
\usepackage{mathptmx}
\usepackage{etoolbox}

\usepackage{bbm} 

\makeatletter
\def\@email#1#2{%
 \endgroup
 \patchcmd{\titleblock@produce}
  {\frontmatter@RRAPformat}
  {\frontmatter@RRAPformat{\produce@RRAP{*#1\href{mailto:#2}{#2}}}\frontmatter@RRAPformat}
  {}{}
}%
\makeatother
\begin{document}

\preprint{AIP/123-QED}

\title[Eigenvalues of regular Hall-plates]{A short note on the eigenvalues of regular Hall-plates}
\author{Udo Ausserlechner}
\email{udo.ausserlechner@infineon.com}
\affiliation{Infineon Technologies Austria AG, Siemensstrasse 2, A-9500 Villach, Austria.
ORCID ID: https://orcid.org/0000-0002-8229-9143 }


\date{\today}

\begin{abstract}
This note is about uniform, plane, singly connected, regular Hall-plates with an arbitrary number of contacts exposed to a uniform magnetic field of arbitrary strength. In practice, the regular symmetry is the most common one. If the Hall-plates are mapped conformally to the unit disk, regular symmetry means that all contacts are equally large and all contact spacings are equally large, yet the contact spacing may be different from the size of the contacts. The indefinite conductance matrices of such Hall-plates are circulant matrices, whose complex eigenvalues are computed in closed form. 
\end{abstract}

\maketitle

\section{Definitions}
\label{sec:fundamentals}

In electrically linear Hall-plates the currents into the terminals are linear combinations of the potentials at the terminals. Suppose that the Hall-plate has $N$ terminals. We group all $N$ currents to a vector $\bm{I}$ and all $N$ voltages to a vector $\bm{V}$. Then we express the linear combination as a matrix multiplication $\bm{I} = {^i}\!\bm{G}\bm{V}$, with the \emph{indefinite} conductance matrix ${^i}\!\bm{G}$. Thereby, the reference potential (ground node) is arbitrary -- none of the terminals needs to be grounded. If we ground the $\ell$-th terminal, we write $\bm{I}=\bm{G}\bm{V}$, where we delete the $\ell$-th current and voltage in $\bm{I},\bm{V}$, respectively, and we delete the $\ell$-th row and column in ${^i}\!\bm{G}$ to get $\bm{G}$. The indefinite conductance matrix is positive semi-definite -- its determinant vanishes and therefore it cannot be inverted. The definite conductance matrix of any passive (= dissipative) system is positive definite, its determinant is positive, and an inverse exists: $\bm{V}=\bm{R}\bm{I}$ with the resistance matrix $\bm{R}=\bm{G}^{-1}$. 

In the presence of the Hall-effect Ohm's law is 
\begin{equation}\label{eq:RMFoCD2}
\bm{E}=\rho\bm{J}-\rho\mu_H\bm{J}\times (B_z\bm{n}_z) , 
\end{equation}
with the electric field $\bm{E}$, the current density $\bm{J}$, the unit vector $\bm{n}_z$ orthogonal to the plane Hall-plate, with the $z$-component of the magnetic flux density $B_z$, and with the Hall-mobility $\mu_H$. Thus, $\bm{E}$ and $\bm{J}$ are not colinear -- there is the Hall-angle $\theta_H$ in-between, with $\tan(\theta_H)=\mu_H B_z$. 
We use the following abbreviations, 
\begin{equation}\label{eq:RMFoCD3}
R_\mathrm{sheet} = \frac{\rho}{t_H}, \quad R_\mathrm{sq} = \frac{R_\mathrm{sheet}}{\cos(\theta_H)} ,
\end{equation}
whereby $t_H$ is the thickness of the Hall-plate. If we have a square plate with two contacts fully covering two opposite edges, the resistance between these contacts is the sheet resistance $R_\mathrm{sheet}$ at zero magnetic field, and it is the square resistance $R_\mathrm{sq}$ in the presence of a magnetic field \cite{Lippmann1958}.

\section{Regular Hall-plates}
\label{sec:regular-sym}

\begin{figure}[t]
  \centering
                \includegraphics[width=0.34\textwidth]{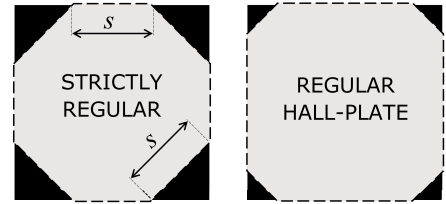}
    \caption{ Examples of strictly regular and regular Hall-plates with four contacts. }
   \label{fig:regular-vs-strictly-regular-Hall-plates}
\end{figure}

Suppose we have a regular boundary with $N$ contacts of equal size, the spacings between the contacts are all equal, yet they are not necessarily identical to the size of the contacts (see Fig. \ref{fig:regular-vs-strictly-regular-Hall-plates}). Suppose we connect an external circuit to this Hall-plate. If we keep the Hall-plate fixed and move the external circuit by one terminal, say clock-wise, then the resulting potentials and currents into the external circuit must remain the same, because the Hall-plate is regular symmetric in shape and the applied magnetic field is parallel to the rotation axis.This means that the entries in the current and voltage vectors move one instance downwards, and the entries in the indefinite conductance matrix move one row down and one column to the right. Both the original and the new ${^i}\!\bm{G}$ matrices must be identical. We can repeat this procedure $N$ times. The conclusion is that the elements along diagonals parallel to the main diagonal are identical, which gives the following patterns for ${^i}\!\bm{G}$ and $\bm{G}$, 
\begin{equation}\label{G-symmetry}
{^i}\!\bm{G} = \left(\!\!\begin{array}{ccccc} A&B&C&D&E\\ E&A&B&C&D\\ D&E&A&B&C\\ C&D&E&A&B\\ B&C&D&E&A \end{array} \!\!\right) \,\Rightarrow\,
\bm{G} = \left(\!\!\begin{array}{cccc} A&B&C&D\\ E&A&B&C\\ D&E&A&B\\ C&D&E&A \end{array} \!\!\right) .
\end{equation}
${^i}\!\bm{G}$ is an $N\times N$ matrix, $\bm{G}$ is an $(N-1)\times (N-1)$ matrix. We get $\bm{G}$ from ${^i}\!\bm{G}$ by deleting the row and column which corresponds to the grounded contact. $\bm{G}$ and ${^i}\!\bm{G}$ are Toeplitz matrices, which have the property that $G_{\ell ,m}=g_{\ell-m}$ -- it depends only on $\ell-m$. Moreover, ${^i}\!\bm{G}$ is a circulant matrix with $g_{\ell-m}=g_{\ell-m+N}$. Third, the sum of entries per row in any indefinite conductance matrix has to vanish, because no currents flow if all contacts are at identical non-zero potential.

If we map a regular Hall-plate to the unit disk, its contact $C_m$ ranges from azimuthal angle 
\begin{equation}\label{eq:RMFoCD100b}
\alpha_m=2\pi \frac{m-1}{N} \;\text{ to }\; \beta_m=\alpha_m+\chi \frac{\pi}{N} , 
\end{equation}
for $\chi\in\left( 0,2\right)$ and $m=1,2,\ldots,N$. Contacts are small for $\chi\to 0$, whereas contact spacings are small for $\chi\to 2$. The case $\chi = 1$ gives \emph{strictly regular} Hall-plates, whose contacts are equally large as the contact spacings (see Fig. \ref{fig:regular-vs-strictly-regular-Hall-plates}). 

We can write down the eigenvalues and eigenvectors of a real-valued circulant matrix \cite{GrayCirculant} 
\begin{equation}\label{eq:RMFoCD52}\begin{split} 
& \bm{{^i}\!G} = \bm{Q} \;\bm{{^i}\!\Gamma} \bm{Q}^C \quad\text{with } \bm{{^i}\!\Gamma} = \mathrm{diag}({^i}\!\gamma_1,{^i}\!\gamma_2,\ldots ,{^i}\!\gamma_N) ,\\
& (\bm{Q})_{k,\ell} = \frac{1}{\sqrt{N}} \exp\left(\frac{2\pi \mathbbm{i} k \ell }{N}\right)  \;\forall k,\ell\in\{1,2,\ldots,N\} .
\end{split}\end{equation} 
$\bm{Q}^C$ is the conjugate complex of $\bm{Q}$, and $\mathbbm{i}=\sqrt{-1}$. $\bm{{^i}\!\Gamma}$ is a diagonal matrix, which comprises all eigenvalues ${^i}\!\gamma_\ell$ of $\bm{{^i}\!G}$. It holds $\bm{Q}=\bm{Q}^T$ and $\bm{Q}\bm{Q}^C=\bm{Q}^C\bm{Q}=\bm{1}$. The eigenvectors of $\bm{{^i}\!G}$ are the columns of $\bm{Q}$, which are coefficient vectors of the discrete Fourier transform (DFT). It generally holds 
\begin{equation}\label{eq:RMFoCD69a}
{^i}\!\gamma_{N-\ell}={^i}\!\gamma_\ell^C ,
\end{equation} 
for the eigenvalues of any real circulant matrix, where $z^C$ is the conjugate complex of $z$.

\section{How to compute the eigenvalues of $\bm{{^i}\!G}$} 
\label{sec:iG}

We use Ref.~ \onlinecite{Homentcovschi2019}. Their equation (17) reads in our notation
\begin{equation}\label{eq:RMFoCD100}
\sum_{m=1}^N B_{k,m} \phi_m = \sum_{m=1}^N  \frac{-A_{k,m} \psi_m}{\cos(\theta_H)} \text{ for }k\in\{1,\ldots,N-1\} ,
\end{equation} 
whereby $\phi_m$ is the potential on $C_m$ (= the $m$-th peripheral contact), and $\psi_m$ is the stream function on the insulating peripheral segment between $C_m$ and $C_{m+1}$, with $C_N=C_0$. 
In Ref.~\onlinecite{Homentcovschi2019} the matrices $\bm{A},\bm{B}$ are defined as
\begin{equation}\label{eq:RMFoCD101}\begin{split}
& A_{k,m} = \int_{\beta_m}^{\alpha_{m+1}}\frac{h(\tau)\,\mathrm{d}\tau}{\sin((\tau-\beta_k)/2)\sin((\tau-\beta_N)/2)} , \\
& B_{k,m} = \int_{\alpha_m}^{\beta_m}\frac{h(\tau)\,\mathrm{d}\tau}{\sin((\tau-\beta_k)/2)\sin((\tau-\beta_N)/2)}, \\
&\text{with } h(\tau) = \prod_{\ell=1}^N \left| \frac{\sin((\tau-\beta_\ell)/2)}{\sin((\tau-\alpha_\ell)/2)} \right|^{(1/2+\theta_H/\pi)} .
\end{split}\end{equation} 
Note that in (\ref{eq:RMFoCD100}) the sums extend over $N$ terms, whereas in Ref.~\onlinecite{Homentcovschi2019} they had only $N-1$ terms. Hence, in contrast to Ref.~\onlinecite{Homentcovschi2019}, we do not require $\phi_N=0$ and $\psi_N=0$. This enables us to work with the indefinite conductance matrix, which is circulant, and therefore it lends for a simpler mathematical treatment than the definite conductance matrix. Therefore, our matrices $\bm{A},\bm{B}$ have $N$ columns and $N-1$ rows. Equation (\ref{eq:RMFoCD100}) must hold, if we add an arbitrary constant to all $\phi_k$ or to all $\psi_k$. Consequently it must hold 
\begin{equation}\label{eq:RMFoCD102}
\sum_{m=1}^N A_{k,m} = 0 \text{ and } \sum_{m=1}^N B_{k,m} = 0 \text{ for }k\in\{1,\ldots,N-1\} .
\end{equation}
Multiplying the left equation of (\ref{eq:RMFoCD102}) with $\psi_N/\cos(\theta_H)$ and adding it to (\ref{eq:RMFoCD100}) gives 
\begin{equation}\label{eq:RMFoCD103}
\sum_{m=1}^N B_{k,m} \phi_m = \sum_{m=1}^N -A_{k,m} \frac{\psi_m-\psi_N}{\cos(\theta_H)} \text{ for }k\in\{1,\ldots,N-1\} .
\end{equation} 
The currents and the stream function are linked via (8) in Ref.~\onlinecite{Homentcovschi2019} (see also (18) in  Ref.~\onlinecite{Ausserlechner2019b}), 
\begin{equation}\label{eq:RMFoCD104}
I_k = \frac{\psi_k-\psi_{k-1}}{R_\mathrm{sheet}} .
\end{equation} 
This means 
\begin{equation}\label{eq:RMFoCD105}\begin{split}
& \left.\begin{array}{rcl} I_1&=&(\psi_1-\psi_N)/R_\mathrm{sheet}\\ I_1+I_2&=&(\psi_2-\psi_N)/R_\mathrm{sheet}\\ & \vdots & \\ I_1+\ldots I_{N-1}&=&(\psi_{N-1}-\psi_N)/R_\mathrm{sheet}\\ I_N&=&0 \end{array} \right\} \\
&\qquad\quad \Rightarrow \bm{L_1}\bm{I} = \frac{1}{R_\mathrm{sheet}}\left(\begin{array}{c}\psi_1-\psi_N\\ \psi_2-\psi_N\\ \vdots\\ \psi_{N-1}-\psi_N\\0 \end{array}\right) , \\
& \qquad\qquad\text{with } \bm{L_1} = 
 \left(\! {\begin{array}{ccccc} 1 & 0 & 0 & \cdots & 0 \\ 1 & 1 & 0 & \cdots & 0 \\ 1 & 1 & 1 & \cdots & 0 \\ \vdots & \vdots & \vdots &\ddots & \vdots \\ 1 & 1 & 1 & \cdots & 1 \\ \end{array} }\!\right) .
\end{split}\end{equation} 
Inserting (\ref{eq:RMFoCD105}) into (\ref{eq:RMFoCD103}) gives in matrix form 
\begin{equation}\label{eq:RMFoCD106}
\bm{B} \bm{\phi} = -R_\mathrm{sq}\bm{A}\bm{L_1}\bm{I} = -R_\mathrm{sq}\bm{A}\bm{L_1}\bm{{^i}\!G}\bm{\phi} .
\end{equation} 
This equation is valid for arbitrary voltage vectors $\bm{\phi}$, therefore we can skip $\bm{\phi}$ on both sides of (\ref{eq:RMFoCD106}). 
Next we introduce the matrix 
\begin{equation}\label{eq:RMFoCD8}\begin{split}
& \bm{\hat{1}} = 
 \left(\! {\begin{array}{ccccc} 0 & 1 & 0 & \cdots & 0 \\ 0 & 0 & 1 & \cdots & 0 \\ \vdots & \vdots & \vdots &\ddots & \vdots \\ 0 & 0 & 0 & \cdots & 1 \\ 1 & 0 & 0 & \cdots & 0 \\ \end{array} }\!\right) .
\end{split}\end{equation}
$\bm{\hat{1}}$ is a circulant $N\times N$ matrix, which is obtained from the unit matrix $\bm{1}$ by shifting all entries up once, in a rolling way. It holds $\bm{\hat{1}}^{-1}=\bm{\hat{1}}^T$, $\mathrm{det}(\bm{1}-\bm{\hat{1}})=0$, and $\mathrm{det}(\bm{1}-\bm{\hat{1}}^T)=0$.
Multiplying (\ref{eq:RMFoCD106}) from right with $(\bm{1}\!-\!\bm{\hat{1}}^T\!)$ gives 
\begin{equation}\label{eq:RMFoCD107}\begin{split}
\bm{B} (\bm{1}\!-\!\bm{\hat{1}}^T\!) & = -R_\mathrm{sq}\bm{A}\bm{L_1}\bm{{^i}\!G}(\bm{1}\!-\!\bm{\hat{1}}^T\!) \\ 
& = -R_\mathrm{sq}\bm{A}\bm{L_1}(\bm{1}\!-\!\bm{\hat{1}}^T\!)\bm{{^i}\!G} ,
\end{split}\end{equation} 
whereby we used the fact that $\bm{{^i}\!G}$ and $(\bm{1}\!-\!\bm{\hat{1}}^T\!)$ are circulant matrices, and therefore they commute. Writing out the product in detail gives $\bm{A}\bm{L_1}(\bm{1}-\bm{\hat{1}}^T) = $ 
\begin{equation}\label{eq:RMFoCD108}\begin{split}
& \bm{A} \left(\! {\begin{array}{ccccc} 1 & 0 & \cdots & 0 \\ 1 & 1 & \ddots & 0 \\ \vdots & \vdots &\ddots & \ddots \\ 1 & 1 & \cdots & 1 \\ \end{array} }\!\right)
\left(\!\begin{array}{ccccc} 1&0&\cdots &0&-1\\ -1&1&\ddots &0&0\\0&-1&\ddots &0 &0\\ \vdots & \ddots & \ddots & 1 & 0 \\ 0&0&\ddots &-1&1 \end{array}\!\right) \\
& = \bm{A}\left(\!\begin{array}{ccccc} 1&0&\cdots &0&-1\\ 0&1&\ddots &0&-1\\0&\ddots &\ddots &\ddots &\vdots \\ 0 & \ddots & \ddots & 1 & -1 \\ 0&0&\cdots &0&0 \end{array}\!\right) = \bm{A} ,
\end{split}\end{equation} 
whereby we used (\ref{eq:RMFoCD102}) in the last equality of (\ref{eq:RMFoCD108}). Equation (\ref{eq:RMFoCD108}) is remarkable, because $\bm{A}\bm{L_1}(\bm{1}-\bm{\hat{1}}^T) = \bm{A}$ even though $\bm{L_1}(\bm{1}-\bm{\hat{1}}^T)\ne\bm{1}$.  Inserting (\ref{eq:RMFoCD108}) into (\ref{eq:RMFoCD107}) gives 
\begin{equation}\label{eq:RMFoCD109}\begin{split}
\bm{B} (\bm{1}\!-\!\bm{\hat{1}}^T\!) & = -R_\mathrm{sq}\bm{A}\bm{{^i}\!G} \\ 
\text{with (\ref{eq:RMFoCD52}) }\;\Rightarrow \bm{B} (\bm{1}\!-\!\bm{\hat{1}}^T\!)\bm{Q} & = -R_\mathrm{sq}\bm{A}\bm{Q} \,\bm{{^i}\!\Gamma} . 
\end{split}\end{equation} 
It holds 
\begin{equation}\label{eq:RMFoCD110}\begin{split}
& (\bm{B} \bm{\hat{1}}^T \bm{Q})_{j,m} = \sum_{k=1}^N \sum_{\ell=1}^N B_{j,k} \delta_{k,1+(\ell\,\mathrm{mod}\,N)} Q_{\ell ,m} \\ 
& = \sum_{\ell=1}^N B_{j,1+(\ell\,\mathrm{mod}\,N)} Q_{\ell ,m} =  \sum_{k=0}^{N-1} B_{j,k+1} Q_{k,m} . 
\end{split}\end{equation} 
Inserting this into (\ref{eq:RMFoCD109}) gives 
\begin{equation}\label{eq:RMFoCD111}\begin{split}
{^i}\!\gamma_m R_\mathrm{sq} \sum_{k=1}^N A_{j,k} Q_{k,m} = -\sum_{k=1}^N B_{j,k} ( Q_{k,m}-Q_{k-1,m} ) .
\end{split}\end{equation} 
Inserting (\ref{eq:RMFoCD100b}) into (\ref{eq:RMFoCD101}) and using the product formula \cite{RyshikGradstein} gives 
\begin{equation}\label{eq:RMFoCD112} 
h(\tau) = \sin\left(\frac{\chi\pi}{2}\right) \left| \cot\left(\frac{\chi\pi}{2}\right) - \cot\left(\frac{N\tau}{2}\right) \right|^{(1/2+\theta_H/\pi)} .
\end{equation} 
Back-inserting this into (\ref{eq:RMFoCD101}) gives after some manipulation 
\begin{equation}\label{eq:RMFoCD113}\begin{split}
& A_{k,m} = \frac{-2}{N} (2-\chi) \sin\left(\frac{\chi\pi}{2}\right) \\
&\quad\quad\quad\times\! \int_0^{\pi/2} \!\frac{\left| \cot\left(\frac{\chi\pi}{2}\right) - \cot\left(\frac{\chi\pi}{2}+z(2-\chi)\right) \right|^{(1/2+\theta_H/\pi)}}{\sin((2\!-\!\chi)z/N\!+\!\pi m/N)} \\ 
&\qquad\qquad\qquad\times \frac{\,\mathrm{d}z}{\sin((2\!-\!\chi)z/N\!+\!\pi (m\!-\!k)/N)}, 
\end{split}\end{equation} 
and 
\begin{equation}\label{eq:RMFoCD113b}\begin{split}
&\quad B_{k,m} = \frac{-2}{N}\chi \sin\left(\frac{\chi\pi}{2}\right) \\
&\quad\quad\quad\times\! \int_0^{\pi/2} \!\frac{\left| \cot\left(\frac{\chi\pi}{2}\right) - \cot\left(\chi z\right) \right|^{(1/2+\theta_H/\pi)}}{\sin(\chi (2z\!-\!\pi)/(2N)\!+\!\pi m/N)} \\
&\qquad\qquad\qquad\times \frac{\,\mathrm{d}z}{\sin(\chi (2z\!-\!\pi)/(2N)\!+\!\pi (m\!-\!k)/N)} .
\end{split}\end{equation} 
We insert (\ref{eq:RMFoCD113}), (\ref{eq:RMFoCD113}) into (\ref{eq:RMFoCD111}) and use the following identity (see Appendix) 
\begin{equation}\label{eq:RMFoCD125}\begin{split}
& \sum_{k=1}^N \frac{\exp\left(2\pi \mathbbm{i} k m / N\right)}{\sin(x/N\!+\!\pi k/N)\sin(x/N\!+\!\pi (k\!-\!j)/N)} \\
& = 2N\frac{\sin\left(\pi j m / N\right)}{\sin\left(\pi j / N\right)} \left(\mathbbm{i}\cot(x)\!-\!1\right)\exp\left(\mathbbm{i}m \frac{j\pi\!-\!2x}{N}\right) .
\end{split}\end{equation} 
Finally, this gives the complex eigenvalues ${^i}\gamma_m, m\in\{1,2,\ldots, N\}$ of the indefinite conductance matrix $\bm{{^i}\!G}$ of a regular symmetric Hall-plate with $N$ contacts as a function of the Hall-angle $\theta_H$ and of the contact size $\chi$, 
\begin{widetext}
\begin{equation}\label{eq:RMFoCD1500}\begin{split}
{^i}\gamma_m = & \frac{-2\chi\mathbbm{i}}{R_\mathrm{sq}(2-\chi)}\exp\left(-\mathbbm{i}\pi\frac{m}{N}\right) \sin\left(\pi\frac{m}{N}\right) \\ 
&\times\frac{\int_0^1 \left(1+\mathbbm{i}\cot\left(\frac{\pi}{2}\chi\tau\right)\right) \exp\left(\mathbbm{i}\pi\chi\frac{m}{N}\tau\right) \left[ \cot\left(\frac{\pi}{2}\chi (1-\tau)\right) -\cot\left(\frac{\pi}{2}\chi\right) \right]^{(1/2+\theta_H/\pi)}\,\mathrm{d}\tau}{\int_0^1 \left(1-\mathbbm{i}\cot\left(\frac{\pi}{2}(2-\chi)\tau\right)\right) \exp\left(-\mathbbm{i}\pi(2-\chi)\frac{m}{N}\tau\right) \left[ \cot\left(\frac{\pi}{2}\chi\right) -\cot\left(\frac{\pi}{2}\left(\chi+\tau(2-\chi)\right)\right) \right]^{(1/2+\theta_H/\pi)}\,\mathrm{d}\tau} .
\end{split}\end{equation}
\end{widetext}
The properties of these eigenvalues will be discussed elsewhere. They can be used in (\ref{eq:RMFoCD52}) to compute the conductance matrix, which gives all electrical properties of the Hall-plate (impedances, response to magnetic field, thermal noise, signal-to-thermal-noise ratio over power).

\appendix*

\section{How to compute the sum in (\ref{eq:RMFoCD125})}
\label{sec:sum}

\begin{figure}[t]
  \centering
                \includegraphics[width=0.20\textwidth]{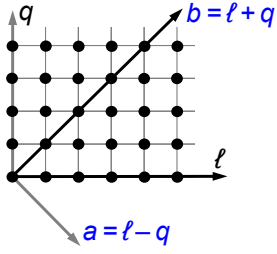}
    \caption{Transformation of indices for a double sum. }
   \label{fig:sum-index-transformation}
\end{figure}

We compute the sum 
\begin{equation}\label{eq:sum1}\begin{split}
\sum_{k=1}^N \frac{\exp\left(2\pi \mathbbm{i} k m / N\right)}{\sin(x/N\!+\!\pi k/N)\sin(x/N\!+\!\pi (k\!-\!j)/N)} ,
\end{split}\end{equation} 
for $N\ge 2$, $j\in\{1,2,\ldots,N-1\}$, and integer $m$.
First we replace $k=N$ by $k=0$, then we express the sines by complex exponentials, 
\begin{equation}\label{eq:sum2}\begin{split}
& \sum_{k=0}^{N-1} \frac{(2\mathbbm{i})^2 \exp\left(2\pi \mathbbm{i} k m / N\right)}{\exp\left(\mathbbm{i}(x/N\!+\!\pi k/N)\right) \exp\left(\mathbbm{i}(x/N\!+\!\pi (k\!-\!j)/N)\right)} \\
&\qquad \times \left[1\!-\!\exp\left(-2\mathbbm{i}\frac{x\!+\!\pi k}{N}\right)\right]^{-1} \\
&\qquad \times \left[1\!-\!\exp\left(-2\mathbbm{i}\frac{x\!+\!\pi (k\!-\!j)}{N}\right)\right]^{-1} .
\end{split}\end{equation} 
Then we make twice a Taylor series expansion of the type $[1-z]^{-1}=1+z+z^2+z^3+\ldots$. This gives 
\begin{equation}\label{eq:sum3}\begin{split}
& \sum_{\ell=0}^\infty\sum_{q=0}^\infty -4\exp\left(-2\mathbbm{i}x\frac{1+\ell+q}{N}\right) \exp\left(\mathbbm{i}\pi j\frac{1+2q}{N}\right) \\
& \qquad\quad \times \sum_{k=0}^{N-1} \exp\left(2\pi \mathbbm{i}k\frac{m-1-\ell-q}{N}\right) .
\end{split}\end{equation} 
The sum over index $k$ vanishes, except for $m-1-\ell-q=-p N$ with $p=0,1,2\ldots$, hence it equals $N\delta_{q,m-1-\ell+pN}$. We re-arrange the double sum over indices $\ell, q$ as a double sum over indices $a,b$ with $a=\ell-q, b=\ell+q$ (see Fig. \ref{fig:sum-index-transformation}). With $b=m-1+p N$ it holds 
\begin{equation}\label{eq:sum4}\begin{split}
& \sum_{p=0}^\infty -4 N \exp\left[\mathbbm{i}\left(\pi j\!-\!2x\right)\frac{m\!+\!pN}{N}\right] \\
& \qquad \times \sum_{a=-(m-1+pN),\Delta a=2}^{m-1+pN} \!\exp\left(\frac{-\mathbbm{i}\pi j a}{N}\right) ,
\end{split}\end{equation} 
whereby the index $a$ is incremented by $\Delta a=2$. We replace the index $a$ by $c$ via $a=2c-m+1-pN$. This gives 
\begin{equation}\label{eq:sum5}\begin{split}
& \sum_{a=-(m-1+pN),\Delta a=2}^{m-1+pN} \exp\left(\frac{-\mathbbm{i}\pi j a}{N}\right) \\ 
& = \exp\left(\mathbbm{i}\pi j\frac{m-1+pN}{N}\right) \sum_{c=0}^{m-1+pN} \exp\left(-2\pi\mathbbm{i}\frac{j}{N}c\right) \\ 
& = (-1)^{j p}\frac{\sin\left(\pi j m/N\right)} {\sin\left(\pi j/N\right)} .
\end{split}\end{equation} 
Inserting(\ref{eq:sum5}) into (\ref{eq:sum4}) and computing the infinite sum over index $p$ finally gives
\begin{equation}\label{eq:sum6}\begin{split}
2N\frac{\sin\left(\pi j m / N\right)}{\sin\left(\pi j / N\right)} \left(\mathbbm{i}\cot(x)-1\right)\exp\left(\mathbbm{i} m \frac{j\pi-2x}{N}\right) .
\end{split}\end{equation}

\bibliography{aipsamp}

\end{document}